\documentstyle[psbox]{WORKSHOP}
\begin{document}

\title{
The RXTE All Sky Monitor: First Year of Performance \\ 
}

\author{
Ronald A. Remillard$^1$ and Alan M. Levine$^1$ \\[12pt]  
%
$^1$ Center for Space Research, Massachusetts Institute of Technology, Cambridge MA 02139 USA \\
{\it E-mail: rr@space.mit.edu} 
}

\abst{ 
The RXTE All Sky Monitor provides a public database
that includes more than one year of X-ray monitoring observations
(2-12 keV) of X-ray binaries and a few active galactic nuclei. The
instrument operates with a 40\% duty cycle, and the exposures
yield roughly 5 celestial scans per day.  There have been 109
source detections, including 16 X-ray transients, the majority of
which are recurrent cases.  The two sources of relativistic radio jets
have exhibited particularly complex light curves and new types
of emission states. Progress has been achieved in understanding the
outburst mechanism via the reported detection of an optical precursor
to the April 1996 X-ray outburst in GRO J1655-40.  The ASM has also
detected state changes in both Cyg X-1 and Cyg X-3, leading to new
constraints on the accretion disk geometry associated with the
``soft/high'' state.  X-ray variations are seen in extragalactic
nuclei, such as NGC4151 and Mkn501, providing new opportunities for
multifrequency timing or spectral studies.  The ASM archive
empowers observers with the opportunity for state-dependent observing
programs with RXTE and other instruments. The ASM also provides a
long-term context for source behavior, and this knowledge may be
crucial in shaping the interpretation of brief observations with other
telescopes.
  }

\kword{surveys --- X-rays: general --- X-rays: stars}

\maketitle

\section{Introduction}

The All Sky Monitor (ASM) on the {\it Rossi X-ray Timing Explorer}
(RXTE) has been regularly observing bright celestial X-ray sources
since 1996 February 22.  Detector problems had been encountered during
the first days of operation (1996 January 5-12), but the instrument
has remained stable under an operation plan restricted to
low-background regions of the RXTE orbit (580 km altitude). In
addition there is an on-board monitor system that switches off high
voltage in the event of moderately high rates in any of several
different measures of the detector background. The ASM currently
operates with 20 of the original 24 detector anodes, and the typical
observation duty cycle is 40\%, with the remainder of the time lost to
the high-background regions of the orbit, spacecraft slews, and
instrument rotation or rewinds. The net yield from the ASM exposures
is about 5 celestial scans per day, excluding regions near the
Sun. All of the ASM calibrations, data archiving, the derivation of
source intensities, and efforts to find new sources are carried out
with integrated efforts of the PI team at M.I.T. and the RXTE Science
Operations Center at Goddard Space Flight Center.

The ASM instrument consists of three scanning shadow cameras (SSC)
attached to a rotating pedestal. Each camera contains a
position-sensitive proportional counter, mounted below a wide-field
collimator that restricts the field of view (FOV) to 6$^{o}$ x
90$^{o}$ FWHM and 12$^{o}$ x 110$^{o}$ FWZI. One camera (SSC3) points
in the same direction as the ASM rotation axis. The other two SSCs are
pointed perpendicular to SSC3.  The latter cameras face the same
direction but the long axes of their collimators are tilted by
+12$^{o}$ and -12$^{o}$, respectively, relative to the ASM rotation
axis. The ASM can be rotated so that the co-pointing SSCs are aligned
with the larger instruments of RXTE (i.e. the PCA and HEXTE). The top
of each collimator is covered with an aluminum plate perforated by 6
parallel (and different) series of narrow, rectangular slits that
function as a coded mask by casting a two-dimensional shadow pattern
for each X-ray source in the collimator's field of view. Further
information on this instrument is given by Levine et al.  (1996).

The ASM raw data includes 3 types of data products that are tabulated
and formatted for telemetry by the two ASM event analyzers in the RXTE
Experiment Data System.  In the current observing mode, position
histograms are accumulated for 90 s ``dwells'' in which the cameras'
FOVs are fixed on the sky. Each dwell is followed by a 6$^{o}$
instrument rotation to observe the adjacent patch of sky.  The
rotation plans for ASM dwell sequences are chosen to avoid having any
portions of the Earth in the FOVs of SSCs 1 and 2. The position
histograms are accumulated in three energy channels: 1.5-3.0, 3-5,
and 5-12 keV. The second data product consists of various
measurements from each camera binned in time. These data are useful in
studying bright X-ray pulsars, bursters, gamma ray bursts, and several
other categories of rapid variability.  The ``good events'' from each
camera are recorded for each energy channel in 0.125 s bins, while 6
different types of background measures are recorded in 1 s bins.
Finally, 64-channel X-ray spectra from each camera are output every 64
s. These data provide a means of monitoring the detector gain, since
we may integrate the spectra over long time scales to observe the 5.9
keV emission line from the weak $^{55}$Fe calibration sources mounted
in each collimator.  In addition, ASM spectra may be useful in
investigations of spectral changes in very bright X-ray sources.

The ASM data archive is a public resource available for both planning
purposes and scientific analysis.  The ``realtime'' and archived
source histories are available in FITS format from the RXTE guest
observer facility:
http://heasarc.gsfc.nasa.gov/docs/xte/xte\_1st.html. The ASM archive
is also available in ASCII table format from the ASM web site at
M.I.T.: http://space.mit.edu/XTE/XTE.html.

\section{ASM Performance and Systematic Issues}

Since this forum addresses many instrumental topics related to all-sky
monitors, we briefly review performance and calibration issues
encountered with the RXTE ASM.  The static model for the position of
the cameras and the ASM rotation axis yields a net uncertainty (rms)
of about 45'', as determined from observations of Sco X-1.  We
minimize the effects of this uncertainty by allowing a camera position
to float during data analysis, using chi square to obtain the best fit
for detected X-ray sources that lie in a camera's FOV. We estimate
that the ASM will obtain positions for new X-ray sources with
uncertainties of 2' to 12' for sources in range of 1 Crab to 30 mCrab,
respectively.

The core of the ASM analysis task is to deconvolve the position
histograms into the X-ray shadow patterns for individual X-ray
sources. Our geometric model for the mask, collimator, and anode
alignments was significantly revised in 1997 Jan, leading to a
complete re-analysis of the ASM database. The primary feature of this
revision was the introduction of time-dependent calibrations of the
physical to ``electronic'' position relationships that governs the
locations of mask shadows in the position histograms. Observations of
Sco X-1 clearly demonstrated variable degrees of evolution of each
anode, appearing as secular shifts in the shadow boundaries in the
position histograms.  The ASM position calibrations are now
periodically revised, and the analysis system computes models for
individual mask shadows by interpolating or extrapolating from these
calibrations for each detector anode.

The observations of the Crab Nebula provide a means of gauging the
magnitude of systematic noise in the ASM light curves.  The observed
variance of the derived intensities is slightly larger than the
estimated statistical variance, implying a systematic uncertainty of
1.9\% of the mean flux. For faint X-ray sources, systematic effects
can be quantified by investigating the light curves for ``blank
field'' positions and quiescent X-ray novae. By binning the ASM source
intensities on very long time scales (e.g. 1 year), there is a +1
mCrab bias in the distribution of mean values. The net effect of systematic
and statistical limitations allows the ASM to reach 5
mCrab at 3 $\sigma$ significance in a typical daily exposure (2-12
keV). All of these uncertainty estimates pertain to celestial
positions well separated from the Galactic center, where source
confusion increases systematic problems.

The observations of the ``polar'' AM Her serves to illustrate the
accuracy achieved in the ASM instrument model. There are no
significant detections of this source with the ASM at 1-day time
scales.  The global average for this source (adjusting for the bias
mentioned above) is only 2.3 mCrab.  However, the analysis of the
individual measurements clearly reveals a periodic signal at 3.0943
$\pm$ 0.0001 hr, which is consistent with the known binary period
(Ritter and Kolb 1995).

\section{Highlights of ASM Results}

Many of the results of ASM science investigations fall under four
categories which are outlined below. Under each topic, we discuss 
representative cases that illustrate the value of monitoring
observations and the effectiveness of the ASM alarm system in
attracting observations at other wavelengths.

\subsection{Light Curves of X-ray Transients}

There has been a remarkable diversity of active X-ray transients
during 1996 and 1997.  The prolonged outbursts from the 2 sources of
relativistic radio jets, GRS1915+105 (X-ray Nova Aql 1992) and
GROJ1655-40 (X-ray Nova Sco 1994), further distinguished these sources
from the typical soft X-ray transients. The ASM light curve for
GRS1915+105 displays the unique behavior of this source, as it wanders
through four emission states that appear to be new variants of the
``very high'' X-ray state of black hole binaries (Morgan, Remillard,
and Greiner 1997; MGR97). Many observatories, including the
revitalized Greenbank Interferometer
(http://info.gb.nrao.edu/gbint/GBINT.html), are now monitoring the
behavior of GRS1915+105. Many groups have begun the investigation of
the complex interrelations between radio emission, IR variations, soft
X-ray instability, and hard X-ray flares in GRS1915+105 as a means to
probe the production of relativistic plasma that powers the radio
flares and jets in this system.

\begin{figure*}[t]
\centering
\psbox[xsize=0.9#1,ysize=0.6#1]
{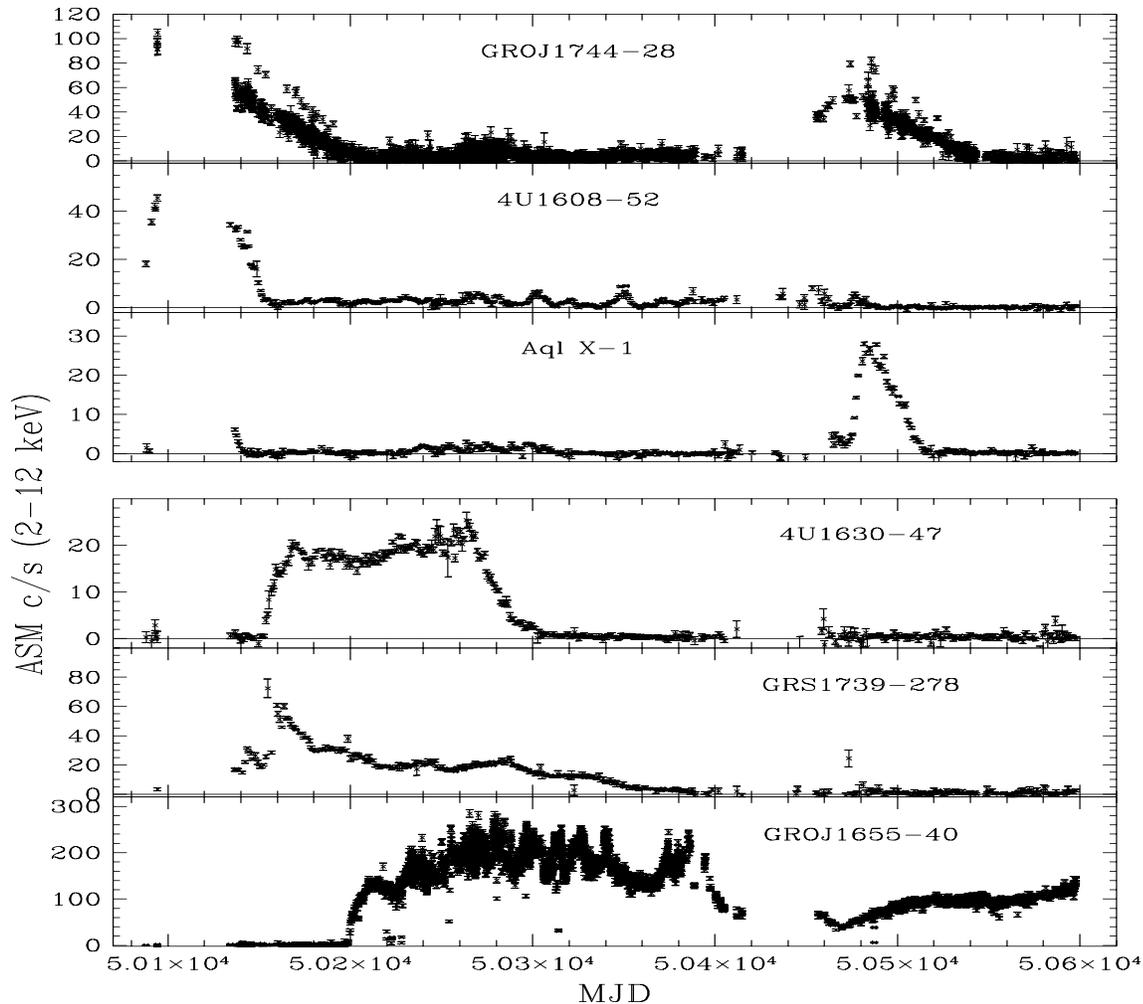}
\caption{ 
ASM light curves for six bright X-ray transients over the time
period 1996 Jan to 1997 May. The top three sources are accreting neutron
stars, while the bottom three are black hole binaries or
candidates. The Crab Nebula produces as ASM rate of 75.5 c/s. }
\end{figure*}

Turning to other transient X-ray sources, most of the two outburst
cycles in GROJ1744-28 were covered with RXTE, as was the prolonged
outburst in a new transient, GRS1739-278.  Major eruptions were also
seen in the recurrent transients 4U1608-52, Aql X-1, and
4U1630-47. The ASM light curves for these 6 sources are shown in
Figure 1. Rapid rise with a brief maximum and an extended period of
low-level emission characterizes both neutron star systems Aql X-1 and
4U1608-52. In contrast, 4U1630-47 rises quickly but then reaches its
luminosity peak very slowly, with an increasingly hard spectrum,
before it decays rapidly (e-fold time of 14 days) with no further
signs of brightness enhancment.  In the case of GRS1739-278, complex
secondary brightening events are seen and the spectrum becomes
increasingly soft with time. All of these ASM results complement
earlier investigations from more limited X-ray monitoring missions
(see Chen, Shrader, and Livio 1997).  The shapes and spectral details
of these outbursts provide substantial challenges for the disk
instability models being applied to soft X-ray transients
(e.g. Cannizzo, Chen, and Livio 1995; Narayan, McClintock, and Yi
1996). Particular support for this model was gained with the detection
of an optical precursor to the April 1996 X-ray ouburst in GROJ1655-40
(Orosz et al. 1997).  The 6-day delay time in the X-ray rise, relative
to optical brightening, has been interpreted in favor of
advection-dominated accretion during the brief period of X-ray
quiescence prior to this event (Hameury et al. 1997).

\begin{figure*}[t]
\centering
\psbox[xsize=0.9#1,ysize=0.6#1]
{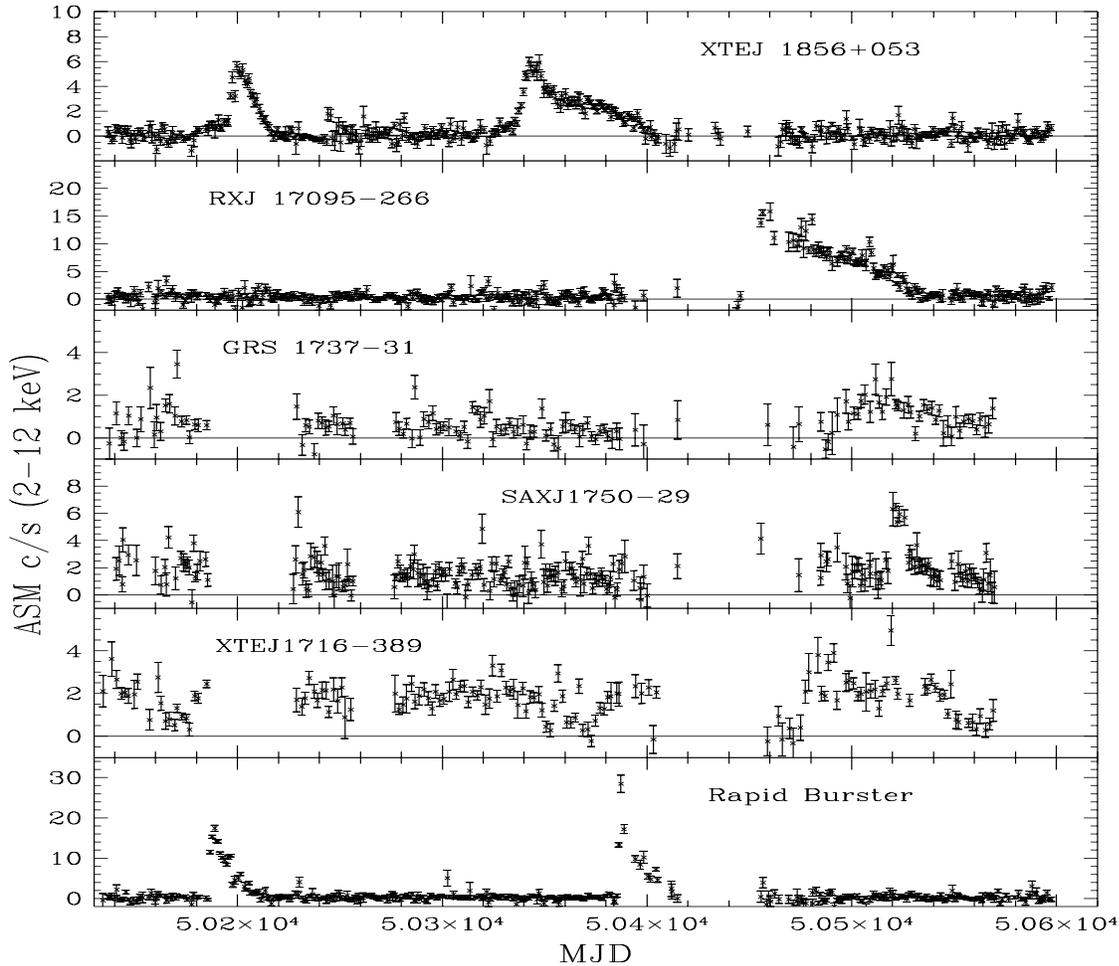}
\caption{ ASM light curves (1996 Feb to 1997 May) for six faint X-ray
transients. Only the Rapid Burster and RXJ17095-266 were known prior
to 1996.  }
\end{figure*}

ASM light curves are also available for a number of fainter X-ray
transients, as shown in Figure 2.  These results are displayed in
1-day or 2-day time bins. Of these 6 cases, only the Rapid Burster and
RXJ1709-266 were known prior to 1996. Again, there is a broad
diversity in the time scales for both X-ray decay and
recurrence. Further examples are needed in order to determine whether
the fainter X-ray transients can be understood as more distant
examples of the same parent populations that produce the bright
transients.

\subsection{Binary Periods and Super-Orbital Periods}

The ASM has already accumulated about two dozen detections of orbital
or ``superorbital'' periods, ranging from the well-known eclipsing
neutron star systems with massive supergiant companions to the more
subtle and less regular periodicities frequently associated with
geometric effects related to the precession of an accretion disk
inclined with respect to the binary plane.  The paper by Corbet in
this conference proceedings is devoted to the theme of orbital and
super-orbital periods detected with the RXTE ASM, and readers are
referred to that work for further information on this topic.

\subsection{State Changes in X-ray Binaries}

The topic of aperiodic variability in X-ray binary systems is very
rich and complex, and a thorough description of the applications for
ASM data is well beyond the scope of this paper. However, one aspect
of this phenomenology, that of state changes in X-ray binaries such as
Cyg X-1, Cyg X-3, and the ``microquasars'', may serve to illustrate
the productivity to be gained by coordinating the observations of
wide-field monitors with those of larger, pointing instruments.

\begin{figure*}[t]
\centering
\psbox[xsize=0.8#1,ysize=0.45#1]
{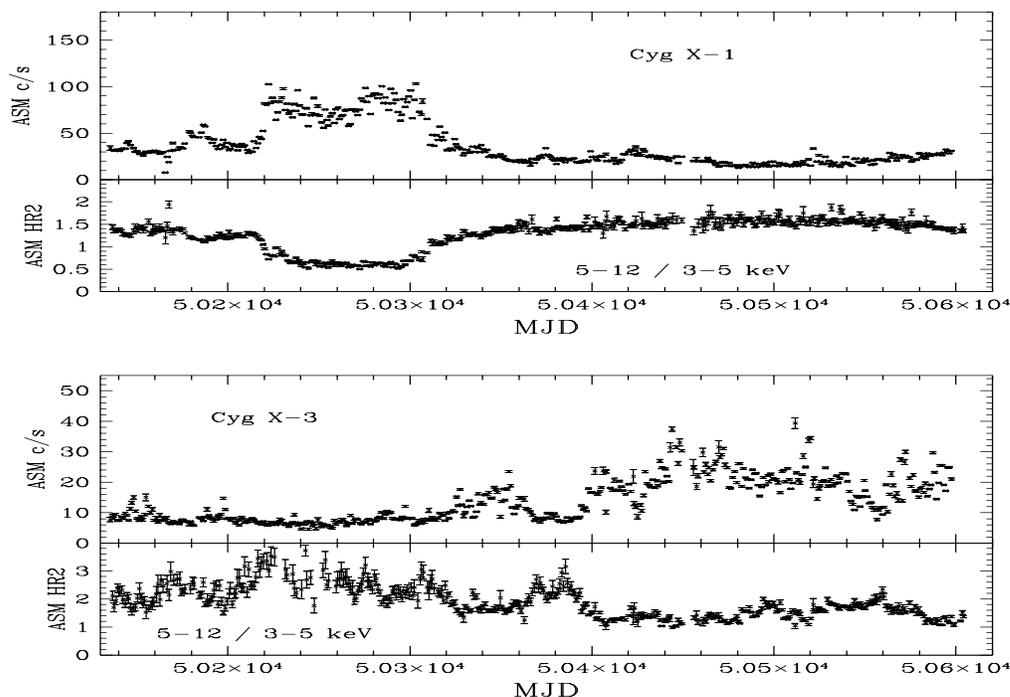}
\caption{ 
ASM light curves and hardness ratios for Cyg X-1 and Cyg X-3 over the time period 1996 Feb to 1997 May. HR2 is the ratio of the flux in the 5-12 keV band to that in the 3-5 keV band. }
\end{figure*}

Both GRS1915+105 and GROJ1655-40 migrate through different X-ray
emission states (e.g. MGR97) on time scales of 3-20 weeks, as
evidenced by correlated changes in the characteristics of the ASM
light curves, the photon spectrum, the shape of the PCA power
spectrum, and the properties of X-ray quasi-periodic oscillations
(QPO). The cause of these transitions remains one of the fundamental
mysteries associated with black hole accretion, the evolution of X-ray
novae, and the mechanism of QPOs. A particularly important aspect of
this science concerns the stationary, high-frequency QPOs seen at 67
Hz in GRS1915+105 (MGR97) and at 300 Hz in GROJ1655-40 (Remillard et
al. 1997). The origin of these QPOs is likley founded in general
relativity, with the QPO frequency dependent on the mass and rotation
of the black hole. While it is not yet clear how to interpret these
high-frequency QPOs, there is strong evidence that their appearance is
correlated with X-ray emission states.  The ASM light curves are
particularly valuable in providing context for these investigations
and in planning future RXTE observations when the X-ray state implies
that these QPOs are likely to recur.

The ASM has also captured the more classical state transitions to the
``soft X-ray high state'' in both Cyg X-1 and Cyg X-3. Figure 3 shows
the ASM light curve and spectral hardness ratio for these sources. In
each case, a 30-day interval with moderate spectral softening preceeds
the main event. In the case of Cyg X-1, the combined coverage of RXTE
and BATSE has demonstrated that there is only a slight (15\%) increase
in total luminosity during the soft/high state (Zhang et al. 1997).
Detailed temporal and spectral analyses of RXTE observations further
indicate that the inner edge of the accretion disk is significantly
closer to the black hole event horizon in the soft/high state, while
the power-law component (associated with inverse Compton emission from
energetic electrons) switches from a flatter spectrum with a thermal
cutoff, in the low/hard state, to a steeper spectrum without a thermal
cutoff from a small emission cloud in the soft/high state (Cui et
al. 1997). The full ramifications of these geometric changes are still
under investigation.

In the case of Cyg X-3, the soft/high transition has also
led to the formation of a radio jet with a velocity that may be as
high as 0.9 c (Ghigo et al. 1997). This high-state episode is still in
progress, and the full story is yet to be told.  Again, the monitoring
efforts from the ASM, BATSE, and the GBI telescope are vital elements
to be combined with observations from the VLA and RXTE pointing
instruments in the effort to understand what has happened to Cyg X-3
during 1997 and what conditions led to the succession of the radio
events that are still underway.

\subsection{Active Galactic Nuclei}

\begin{figure*}[t]
\centering
\psbox[xsize=0.9#1,ysize=0.45#1]
{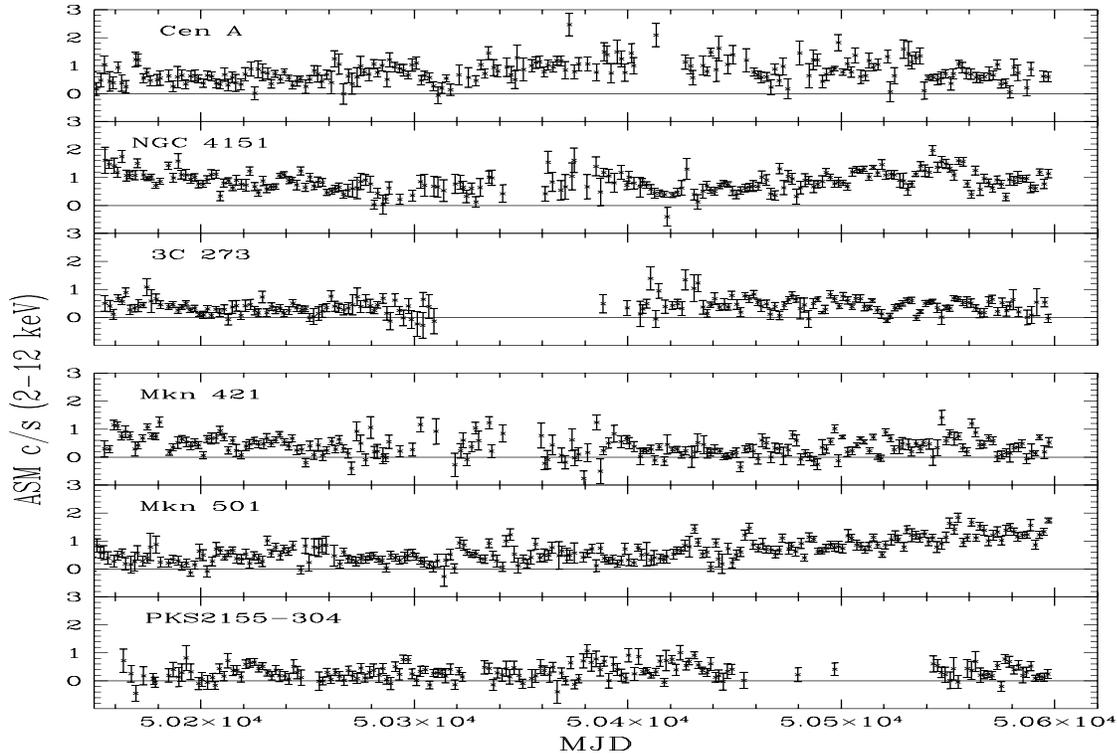}
\caption{ 
ASM light curves of Cen A, an optically obscured AGN, the Seyfert 1.5 galaxy NGC 4151, the quasar 3C273, and 3 BL Lac objects.  The data are displayed in 1-day or 2-day bins. }
\end{figure*}

With a 30 mCrab detection threshold per SSC-dwell, it was not expected
that the RXTE ASM would have the sensitivity necessary to provide
useful information on the different subclasses of active galactic
nuclei (AGN).  There are only a few AGN expected to exceed 3 mCrab at
2-12 keV (e.g. Wood et al. 1984).  However, the ability to combine the
individual ASM measurements into 1-day or 2-day time bins without
encountering significant fluctuations from systematic problems allows
the ASM to monitor many nearby AGN for major changes in X-ray
brightness.  Six examples are shown in Figure 4. Both Cen A and NGC
4151 vary by a factor of two, and the flux from NGC 4151 reaches 24
mCrab at 2-12 keV.  A major outburst in the BL Lac object Mkn 501
began during 1997 January (near MJD 50450), with X-ray intensity
rising from 3 to 20 mCrab. This outburst is correlated with strong
detections of Mkn 501 in TeV $\gamma$-rays with the HEGRA Cherenkov
array (Aharonian et al. 1997). The TeV and X-ray components are very
likley to be inverse-Compton and synchrotron photons emitted by the
same parent population of relativistic electrons. These exciting
results and the opportunities for further detections of AGN outbursts
has led us to increase the total number of ASM-monitored emission-line
AGNs and BL Lac objects from 10 to 74 during 1997 May.

\section*{References}
\re
Aharonian, F. et al. 1997, A\&A, submitted, astro-ph/9706019
\re
Cannizzo, J. K., Chen, W., and Livio M. 1995, ApJ, 454, 880
\re
Chen, W., Shrader, C. R., and Livio, M. 1997, ApJ, in press, astro-ph/9707138
\re
Cui, W., Zhang, S. N., Focke, W., and Swank, J. H. 1997, ApJ, 484, 383
\re
Ghigo, F. D. et al. 1997, Bull. AAS, 29, 841
\re
Hameury, J. M., Lasota, J. P., McClintock, J. E., and Narayan, R. 1997, ApJ, submitted; astro-ph/9703095
\re
Levine, A. M. et al. 1996 ApJL, 469, L33
\re
Morgan E. H., Remillard, R. A. and Greiner, J. 1997, ApJ, 482, 993
\re
Narayan, R., McClintock, J. E., and Yi, I. 1996, ApJ, 457 821
\re
Orosz, J. A., Remillard, R. A., Bailyn, C., and McClintock, J. E. 1997, ApJL, 478, L83 
\re
Remillard, R. A. et al. 1997, Procs. Texas Symposium on Relativistic Astrophysics (1996 December), in press; astro-ph/9705064
\re
Ritter, H, and Kolb, U. 1995, in X-ray Binaries, eds. Lewin and van Paradijs (Cambridge: Cambridge Univ. Press), 578
\re
Wood et al. 1984, ApJS, 56, 507
\re
Zhang, S. N. et al. 1997, ApJL, 477, L95
\re
\label{last}

\end{document}